\documentstyle[prl,twocolumn,aps,english]{revtex}

\def\beq{\begin{equation}}
\def\eeq{\end{equation}}
\def\6{\langle}
\def\9{\rangle}
\def\half{\mbox{$1\over2$}}
\def\pp{{\bf p}}

\begin{document}
\draft

\title{Quantum Entropy and Special Relativity}

\author{Asher Peres, Petra F. Scudo, and Daniel R. Terno}
\address{Department of Physics, Technion---Israel Institute of
Technology, 32000 Haifa, Israel}

\maketitle
\begin{abstract}
We consider a single free spin-\half\ particle. The reduced density 
matrix for its spin is not covariant under Lorentz transformations. 
The spin entropy is not a relativistic scalar and has no invariant
meaning.
\end{abstract}

\bigskip
\pacs{PACS numbers: 03.65.Ta, 03.30.+p.}

The relationship of thermodynamics to relativity theory has been an
intriguing problem for many years \cite{tolman}, and it took a new twist
when quantum properties of black holes were discovered \cite{wald}.
In this Letter, we shall investigate a much simpler problem: the
relativistic properties of spin entropy for a single, free particle of
spin~\half\ and mass $m>0$. We show that the usual definition of quantum
entropy \cite{qt} has no invariant meaning in special relativity.

The reason is that under a Lorentz boost, the spin undergoes a
Wigner rotation \cite{wigner} whose direction and magnitude depend
on the momentum of the particle. Even if the initial state is a
direct product of a function of momentum and a function of spin, the
transformed state is not a direct product. Spin and momentum appear
to be entangled. This is not the familiar type of entanglement which
can be used for quantum communication, because both degrees of freedom
belong to the same particle, not to distinct subsystems that could be
widely separated.

The quantum state of a spin-\half\ particle can be written, in the
momentum representation, as a two-component spinor,

\beq \psi(\pp)={a_1(\pp)\choose a_2(\pp)}, \eeq
where the amplitudes $a_r$ satisfy $\sum_r\int|a_r(\pp)|^2 d\pp=1$. The
normalization of these amplitudes is a matter of convenience, depending
on whether we prefer to include a factor $p_0=\sqrt{m^2+\pp^2}$ in it,
or to have such factors in the transformation law as in Eq.~(\ref{tf})
below \cite{weinberg}. Following Halpern \cite{halpern}, we shall use
the second alternative, because this is the nonrelativistic notation
which appears in the definition of entropy. We use natural units: $c=1$.

Here we emphasize that we consider normalizable states, in the momentum
representation, not momentum eigenstates as usual in textbooks on
particle physics. The latter are chiefly concerned with the computation
of $\6\mbox{in}|\mbox{out}\9$ matrix elements needed to obtain cross
sections and other asymptotic properties. However, in general a particle
has no definite momentum. For example, if an electron is elastically
scattered by some target, the electron state after the scattering is a
superposition that involves momenta in all directions.

In that case, it still is formally possible to ask, in any Lorentz
frame, what is the value of a spin component in a given direction (this
is a legitimate Hermitian operator). We show that the answers to such
questions, asked in different Lorentz frames, are not related by any
transformation group. The purpose of the present work is to make a
first step toward a relativistic extension of quantum information
theory. The important issue does not reside in asymptotic properties,
but how entanglement (a communication resource) is defined by different
observers. Earlier papers on this subject used momentum eigenstates,
just as in particle physics \cite{czachor}. Here we show that radically
new properties arise when we consider localized quantum states.

The density matrix corresponding to Eq.~(\theequation) is

\beq \rho(\pp',\pp'')=\left( \begin{array}{ccc}
  a_1(\pp')a_1(\pp'')^* & & a_1(\pp')a_2(\pp'')^* \\
  a_2(\pp')a_1(\pp'')^* & & a_2(\pp')a_2(\pp'')^* \end{array}\right). \eeq
The reduced density matrix for spin, irrespective of momentum, is
obtained by setting $\pp'=\pp''=\pp$ and integrating over \pp. It can be
written as

\beq \tau=\half\left( \begin{array}{ccc} 
  1+n_z & & n_x-in_y \\ n_x+in_y & & 1-n_z \end{array}\right), \eeq
where the Bloch vector $\bf n$ is given by

\beq n_z= \int(|a_1(\pp)|^2-|a_2(\pp)|^2)\,d\pp, \eeq
and
\beq n_x-in_y=\int a_1(\pp)a_2(\pp)^*d\pp. \eeq

The reduced density matrix $\tau$ gives statistical predictions for
the results of measurements of spin components by an ideal apparatus
which is not affected by the momentum of the particle. The corresponding
entropy is \cite{qt}

\beq S=-{\rm tr\,}(\tau\,\ln\tau)=-\sum\lambda_j\ln\lambda_j,\eeq
where

\beq \lambda_j=(1\pm|{\bf n}|)/2, \eeq
are the eigenvalues of $\tau$.

It is well known that ignoring some degrees of freedom usually
leaves the others in a mixed state. What is not obvious is that the
amount of mixing depends on the Lorentz frame used by the observer.
Indeed consider another observer who moves with a constant velocity
with respect to the one who prepared the above state. In the Lorentz
frame where the second observer is at rest, the same spin-\half\
particle has a state

\beq \phi(\pp)={b_1(\pp)\choose b_2(\pp)}. \eeq
The transformation law is \cite{weinberg,halpern},

\beq b_r(\pp)=[(\Lambda^{-1}p)_0/p_0]^{1/2}\,\sum_s
 D_{rs}[\Lambda,(\Lambda^{-1}p)]\,a_s(\Lambda^{-1}p) \label{tf}, \eeq
where $D_{rs}$ is the Wigner rotation matrix \cite{wigner} for a
Lorentz transformation $\Lambda$ (explicitly given in Ref.
\cite{halpern}, p.~134).

As an example, consider a particle prepared with spin in the $z$
direction, so that in the Lorentz frame of the preparer $a_2=0$. The
Bloch vector has only one component, $n_z=1$, and the spin entropy is
zero. When that particle is described in a Lorentz frame moving with
velocity $\beta$ in the $x$ direction, we have, explicitly

\beq b_1(\pp)=K\,[C\,(q_0+m)+S\,(q_x+iq_y)]\,a_1({\bf q}), \eeq

\beq b_2(\pp)=K\,S\,q_z\,a_1({\bf q}),\eeq
where we have used the following notations: $q_\mu=(\Lambda^{-1}p)_\mu$
is the momentum variable in the original Lorentz frame, $\gamma\equiv
(1-\beta^2)^{-1/2}\equiv\cosh\alpha$,

\beq C\equiv\cosh(\alpha/2), \qquad S\equiv\sinh(\alpha/2), \eeq
and
\beq K\equiv[q_0/p_0(q_0+m)(p_0+m)]^{1/2}.\eeq

The new reduced density matrix $\tau'$ is obtained as before by
integrating over the momenta. Consider in particular the case where
$a_1(\pp)$ is a Gaussian (a minimum uncertainty state):

\beq a_1(\pp)=(2\pi)^{-3/4}w^{3/2}\,\exp(-\pp^2/2w^2).\eeq
All calculations can be done analytically. To leading order of
$w/m\ll1$, we obtain for the new components of the Bloch vector
(defined as above) $n'_x=n'_y=0$, and

\beq n'_z=1-(w\tanh\mbox{$\alpha\over2$}/2m)^2.\eeq
In the new Lorentz frame, the entropy is positive:

\beq S\simeq t(1-\ln t),\eeq
where $t=w^2\tanh^2\mbox{$\alpha\over2$}/8m^2$. (Note that if the
momentum has a sharp value, $w$ and $t$ vanish, and therefore the
entropy also vanishes, as expected.)

The reduced density matrix $\tau$ has no covariant transformation
law, except in the limiting case of sharp momenta (only the {\it
complete\/} density matrix has one). There is an analogous situation
in classical statistical mechanics: a Liouville function can be
defined in any Lorentz frame \cite{balescu}, but it has no definite
transformation law from one frame to another. Only the complete
dynamical system has a transformation law \cite{open}.

It is important to understand how linearity is lost in this purely
quantum mechanical problem. The momenta \pp\ transform linearly,
but the law of transformation of spin components depends explicitly
on~\pp. When we compute $\tau$ by summing over momenta in $\rho$,
all knowledge of these momenta is lost and it is then impossible to
obtain $\tau'$ by transforming $\tau$. Not only linearity is lost, but
the result is not nonlinearity in the usual sense of this term. It is
the absence of {\it any\/} definite transformation law which depends
only on the Lorentz matrix.

Naturally, linearity is still present in a trivial sense. If $\rho=\sum
c_j\rho_j$, then likewise  $\tau=\sum c_j\tau_j$, and after a Lorentz
transformation $\rho'=\sum c_j\rho'_j$, and $\tau'=\sum c_j\tau'_j$.
However, even if we know the values of the coefficients $c_j$, the
mere knowledge of the reduced density matrix $\tau$ is insufficient to
obtain $\tau'$ (although the knowledge of the complete density matrix
$\rho$ does determine $\rho'$).

In the case investigated above, the entropy computed in the moving
frame is larger than the entropy in the original frame, which
was zero. This does not mean that a Lorentz transformation always
increases the entropy: if we have a particle in the state $b_r(\pp)$ as
the one given above, with a positive entropy, then an observer moving
in the $-x$ direction with the appropriate velocity would say that its
state is given by $a_s(\pp)$. For that observer, the entropy is zero.
An invariant definition of entropy could be the
minimal value of the latter, in any Lorentz frame. (Likewise, the mass
of a classical system is defined as the minimal value of its energy,
in any Lorentz frame.) Another possibility would be to use the Lorentz
frame where $\6\pp\9=0$. It is unlikely that such definitions lead to
analytical formulas, although in any particular case the result can
easily be obtained by numerical methods.

An interesting problem is the relativistic meaning of quantum
entanglement when there are several particles. For two particles,
an invariant definition of entanglement would be to compute it the
Lorentz ``rest frame'' where $\6\sum\pp\9=0$. However, this simple
definition is not adequate when there are more than two particles,
because there appears a problem of cluster decomposition: each subset of
particles may have a different rest frame. This is a difficult problem,
which is beyond the scope of this Letter.

In summary, we have shown that the notion ``spin state of a particle''
is meaningless if we don't specify its complete state, including the
momentum variables. It is possible to formally define spin in any
Lorentz frame, but there is no relationship between the observable
expectation values in different Lorentz frames.

\bigskip Work by AP was supported by the Gerard Swope Fund and the
Fund for Encouragement of Research. PFS and DRT were supported by
grants from the Technion Graduate School. Part of this research was done
while DRT was a guest of the Institute for Quantum Information at
Caltech. Special thanks are due to Barbara Terhal, David DiVincenzo, and
Patrick Hayden for stimulating discussions.

\end{document}